\documentstyle[aps,prd,eqsecnum,preprint,axodraw,tighten,floats,epsf]{revtex}

\def\lsim{\stackrel{<}{\sim}}
\def\beq{\begin{equation}}
\def\eeq{\end{equation}}
\def\ba{\begin{array}}
\def\ea{\end{array}}
\def\bea{\begin{eqnarray}}
\def\eea{\end{eqnarray}}
\begin{document}

\preprint{UFIFT-HET-02-10}

\vspace{1cm}

\title{Solar Wakes of Dark Matter Flows}
\author{Pierre Sikivie and Stuart Wick}
\address{
   Department of Physics,
   University of Florida,
   Gainesville, FL,  32611\\
   March 25, 2002}
\maketitle

\begin{abstract} 

We analyze the effect of the Sun's gravitational field on a flow of cold
dark matter (CDM) through the solar system in the limit where the velocity
dispersion of the flow vanishes.  The exact density and velocity
distributions are derived in the case where the Sun is a point mass.  The
results are extended to the more realistic case where the Sun has a finite
size spherically symmetric mass distribution.  We find that regions of
infinite density, called caustics, appear.  One such region is a line
caustic on the axis of symmetry, downstream from the Sun, where the flow
trajectories cross.  Another is a cone-shaped caustic surface near the
trajectories of maximum scattering angle.  The trajectories forming the
conical caustic pass through the Sun's interior and probe the solar
mass distribution, raising the possibility that the solar mass
distribution may some day be measured by a dark matter detector on Earth.  
We generalize our results to the case of flows with continuous velocity
distributions, such as that predicted by the isothermal model of the Milky
Way halo.

\vspace{0.34cm}
\end{abstract}

\newpage

\section{Introduction}
\label{sec:int}

There are compelling reasons to believe that the dark matter of the
universe is constituted in large part of non-baryonic collisionless
particles with very small primordial velocity dispersion \cite{CDM}.  
Generically, such particles are called cold dark matter (CDM).  The
leading CDM candidates are the axion and the lightest supersymmetric
partner in supersymmetric extensions of the Standard Model of elementary
particles. The latter candidate is a weakly interacting massive (mass in
the 10 to 100 GeV range) particle, or WIMP for short.  The former has 
mass in the $10^{-5}$ to $10^{-6}$ eV range and is extremely weakly
interacting, so weakly in fact that it was once thought ``invisible".

Wherever a galaxy forms, the surrounding CDM falls into the galactic
gravitational potential.  Unlike baryons, CDM is collisionless and sloshes
back and forth in the potential well, producing a galactic halo of size
much larger than the extent of the visible matter.  The local (i.e., on
Earth) dark matter density of the Milky Way halo is of order $10^{-24}~
{\rm gr}/{\rm cm^3}$.  This local dark matter is in principle detectable
and considerable effort is being spent on projects devoted to this goal.  
Dark matter axions can be detected by their conversion to microwave
photons in an electromagnetic cavity permeated by a strong static magnetic
field \cite{sik}.  WIMPs can be detected by observing the recoil energy of
nuclei off which WIMPs have scattered elastically \cite{WIMP,rev}.  WIMP
dark matter may also be detected indirectly by observing, on Earth,
neutrinos produced in the annihilation of WIMPs which accumulated in the
interior of the Sun \cite{ind}.

From the point of view of these dark matter searches it is desirable to
know as much as possible about the phase-space distribution $f(\vec{r},
\vec{v})$ of CDM in galactic halos in general, and particularly in the
Milky Way halo. One wants to know not only the local density but also the
local velocity distribution.  In the cavity dark matter axion detector,
the kinetic energy spectrum of dark matter axions can be measured with
great resolution \cite{sik}.  Knowledge of the velocity distribution may
therefore be exploited to increase the sensitivity of the search.  In WIMP
searches, knowledge of the velocity spectrum is necessary to calculate the
annual modulation of the signal \cite{ann}.  This annual modulation is an
important means by which the signal to background ratio can be enhanced.
  
Yet, in spite of this urgent need of the ongoing cold dark matter
searches, there is at present no consensus about the phase-space structure
of galactic halos.  Let us discuss briefly the three main approaches which
have been put forth.

The oldest description of galactic halos is the isothermal model
\cite{iso}. It postulates that the phase-distribution is given by
Boltzmann's law, the energy of each particle being its kinetic energy plus
its gravitational potential energy in the presence of all the other
particles.  It can be shown that the dark matter density then falls off as
$d(r_G) \propto {1 \over r_G^2}$ for large values of the galactocentric
distance $r_G$.  Therefore the model predicts that galactic rotation
curves are flat ($v(r_G) \propto $ constant) at large $r_G$, and that is
indeed what is observed.  Another virtue of the model is that it predicts
that at small $r_G$, within some radius called the core radius $a$, the
density $d$ goes to a constant, and hence that the halo contribution to
galactic rotation curves $v_{\rm halo}(r_G) \propto r_G$ at small $r_G$.  
That also is consistent with observation.  The model predicts all
properties of galactic halos in terms of just two parameters: the velocity
dispersion $\sqrt{\vec{v}^2}$ and the core radius $a$.  In fits to
observed properties of our own halo, $\sqrt{\vec{v}^2}$ is found to be
approximately 270 km/s and $a$ of order several kpc.

The isothermal model is simple and gives an excellent overall description
of what is observed, but there is little justification for its basic
assumption, that the dark matter particles in galactic halos are
thermalized.  Thermalization may have occurred during an early phase of
'violent relaxation' \cite{vio} at the time of collapse of the
protogalaxy.  It is possible that the dark matter that fell onto the
Galaxy at early times was thermalized by that process.  But an isolated
galaxy such as the Milky Way is still accreting dark matter today
because it is surrounded by a sea of dark matter.  (The presence of M31
and the other members of the local group is not relevant for the purpose
of this discussion.) There is no evidence of violent relaxation
occurring in the Galaxy at the present time.  The infalling dark matter
particles form flows which slosh back and forth in the gravitational
potential of the galaxy.  The flows do not get thermalized over the age of
the universe \cite{ips}.  Gravitational scattering by inhomogeneities,
such as globular clusters and molecular clouds, is too weak to diffuse the
flows entirely, except for those flows which always remained within the
inner parts ($r_G \lesssim 20$ kpc, say) of the Galaxy.

The flows can be described in terms of the evolution of the 3-dim.~sheet
on which the dark matter particles lie in 6-dim.~phase-space \cite{ips}.  
Before the onset of galaxy formation, the equation of the sheet is
$\vec{v} = H(t) \vec{r} + \Delta \vec{v}(\vec{r},t)$ where $H(t)$ is the
Hubble expansion rate and $\Delta \vec{v}(\vec{r},t)$ is the peculiar
velocity field associated with density perturbations.  The thickness of
the sheet is the primordial velocity dispersion $\delta v$.  The
primordial velocity dispersion of the leading cold dark matter candidates
is extremely small, of order $10^{-12}c$ for WIMPs and $3\cdot10^{-17}c$
(at most) for axions. The sheet cannot tear and hence its evolution is
constrained by topology.  Where a galaxy forms, the sheet wraps up in
phase-space, turning clockwise in any two dimensional cut $(x, \dot{x})$
of that space.  $x$ is the physical space coordinate in an arbitrary
direction and $\dot{x}$ is the associated velocity.  The outcome of this
process is a discrete set of flows at any point in a galactic halo.  Two
flows are associated with particles falling in and out of the galaxy for
the first time ($n$ = 1), two other flows are associated with particles
falling in and out of the galaxy for the second time ($n$ = 2), and so on.  
The number of such flows at the Earth's location in the Milky Way is of
order 100 \cite{ips}.  The flows of particles which have fallen in and out
of the Galaxy only a small number of times in the past ($n \lsim 20$) are
not expected to be thermalized. The flows with larger $n$ may be
thermalized because they are composed of particles which fell in a 
long time ago.

Caustics appear wherever the projection of the phase-space sheet onto
physical space has a fold \cite{PS1,hog,PS2}.  Caustics are generically
surfaces in physical space.  On one side of the surface there are two more
flows than on the other.  At the surface, the dark matter density is very
large.  It diverges there in the limit of zero velocity dispersion.  
There are two types of caustics in the halos of galaxies, inner and outer.  
The outer caustics are topological spheres surrounding the galaxy.  They
are located near where a given outflow reaches its furthest distance from
the galactic center before falling back in, and are described by $A_2$ (or
'fold')  catastrophes.  The inner caustics are rings \cite{PS1}.  They are
located near where the particles with the most angular momentum in a given
inflow reach their distance of closest approach to the galactic center
before going back out.  A caustic ring is a closed tube whose
cross-section is a $D_{-4}$ (also called {\it elliptic umbilic})
catastrophe \cite{PS2}.  The existence of these caustics and their
topological properties are independent of any assumptions of symmetry.

A second approach to describing the structure of galactic halos is in
terms of these flows and their associated caustics
\cite{ips,stw,PS1,bux,PS2,kin,mw}.  Let us call that approach the 'caustic
ring model' since caustic rings are one of its prominent features.  In the
model the caustic ring radii, and indeed the whole phase-space
distribution, are calculated assuming that halo formation is a
self-similar process \cite{ss,stw}.  The model depends on only three
parameters: the galactic rotation velocity $v_{\rm rot}$, a parameter
$j_{\rm max}$ which is proportional to the amount of angular momentum in
the halo, and a parameter $\epsilon$ which describes the profile of the
initial overdensity from which the galaxy forms.  $\epsilon$ is related to
the slope of the evolved power spectrum of primordial density
perturbations and hence, to the extent that that spectrum is predicted by
CDM cosmology, $\epsilon$ is not a free parameter.  If the primordial
spectrum is Harrison-Zel'dovich, $\epsilon$ is in the range 0.2 to 0.35.  
\cite{stw}. In that range, the self-similar model predicts that the
density $d(r_G) \propto {1 \over r_G^2}$ at large $r_G$. Hence the model
predicts that galactic rotation curves are flat at large $r_G$, which is
consistent with observation.  When $j_{\rm max} = 0$, the $d(r_G) \propto 
{1 \over r_G^2}$ behavior extends to small $r_G$.  The ${1 \over r_G^2}$
singularity at $r_G=0$ is due to the fact that the dark matter particles
move on radial orbits in that limit. However, the dark matter particles
are expected to carry angular momentum.  When $j_{\rm max} \neq 0$, the
density is depleted within an effective core radius $r_c$.  As a result,
the halo contribution $v_{\rm halo}(r_G)$ to the rotation curve goes to
zero as $r_G \rightarrow 0$, which is also consistent with observation.

Evidence for the presence of caustic rings in our galaxy was found in the
distribution of rises in the Milky Way rotation curve, and in a triangular
feature in an IRAS map of the galactic disk \cite{mw}.  The feature is
reminiscent of the shape of the $D_{-4}$ catastrophe.  When the model is
fitted to these observations, its parameters are determined to be:  
$v_{\rm rot} \simeq$ 220 km/s, and $j_{\rm max} \simeq 0.26$ for $\epsilon
= 0.3$.  The model predicts the dark matter flows and caustics in the
solar neighborhood.  Because of our proximity to the fifth caustic ring,
one of the flows dominates over all others combined.  Its density and
velocity vector (known only up to a two-fold ambiguity) relative to the
non-rotating restframe of the Galaxy are \cite{mw}:  \beq d_5^+ = 1.7
\cdot 10^{-24}{{\rm gr} \over {\rm cm}^3},~~ \vec{v}_5 = (470~\hat{\phi}_G
\pm 100~\hat{r}_G){{\rm km} \over {\rm s}}~\ .  \label{bigflow} \eeq
$\hat{r}_G$ is the local unit vector in the direction away from the
galactic center, and $\hat{\phi}_G$ is in the direction of galactic
rotation.  A list of the other flows is given in ref.\cite{bux}.  The
caustic ring model describes the minimum phase-space structure that a
galactic halo must have.  The actual phase-space structure cannot be
simpler than that.

A third approach to studying the structure of galactic halos is by means
of N-body simulations \cite{sim}.  Initial density perturbations are
chosen according to a theoretically expected probability distribution,
e.g. Gaussian density perturbations with the CDM power spectrum evolved
from a flat Harrison-Zel'dovich primordial spectrum.  Overdensities grow
in the simulations and form structures which are identified with galactic
halos.  In the large N limit, the outcome of the simulations can be
identified with the predictions of CDM cosmology.  However, present
constraints on computing power restrict N to be of order $10^6$ per
galactic halo.  In contrast, the number of axions per galactic halo is of
order $10^{84}$, and the number of WIMPs of order $10^{68}$.  Since
phase-space is 6-dim.~, the resolving power of the simulations is only a
factor of ten in each dimension, including the three physical space
dimensions.  The simulations do show that galactic halos are not
thermalized and that they contain discrete flows.  However, the limited
resolution allows only a few flows to be seen.  The simulations predict
that a large galaxy like our own has hundreds of satellites.  The tidal
disruption of these satellites produces low velocity dispersion
streams \cite{sti} which are in effect localized flows.

The purpose of this paper is to analyze the effect of the Sun's
gravitational field on a flow of collisionless dark matter through the
solar system.  This effect was analyzed several years ago by K. Griest 
\cite{gri} in the case of the isothermal model.  He calculated the annual
modulation of the signal in WIMP detectors including the effects both of
the Earth's orbital motion and the Sun's gravity.  Earlier discussions 
of the effect of the Sun's gravity on a flow of ordinary collisionful
matter can be found in papers by H. Bondi and F. Hoyle \cite{bon}, 
J.M.A. Danby and G.L. Camm \cite{dan}, and Danby and T.A. Bray
\cite{bra}.  We will study mainly the effect of the Sun's gravity on a
single flow of collisionless dark matter in the limit of zero velocity
dispersion.  The effect of finite velocity dispersion can be taken account
by integrating over a distribution of initial velocities. Our results can
be applied to the calculation of the annual modulation of the signals in
both the axion and WIMP dark matter detectors.  We plan to do this in
detail in a subsequent paper which will be concerned with the experimental
implications of the present investigation.  In the remainder of this
Introduction, we give a general discussion of the annual modulation of 
the signal in the axion and WIMP detectors, including the role of the 
Sun's gravity.

A flow produces a peak in the energy spectrum of photons produced 
by axion $\rightarrow$ photon conversion in the cavity detector of 
dark matter axions \cite{sik}.  The peak's frequency $\omega$ is
determined by the energy conservation relation,  
$\hbar \omega = m_a (c^2 + {1 \over 2} v_{\rm flow \div lab}^2)$, 
in terms of the flow speed relative to the laboratory.  
Because the Earth rotates, there is a diurnal modulation of the peak's 
frequency.  By measuring the average frequency and the amplitude and 
phase of the modulation, one can derive the velocity vector 
$\vec{v}_{\rm flow \div \oplus}$ of the flow relative to the Earth
($\oplus$).  The height of the peak is proportional to the flow density.  

On top of the diurnal modulation, there is a much larger annual 
modulation because the Earth orbits the Sun ($\odot$):
\beq
\hbar\bar{\omega}(t) = m_a [c^2 + 
{1 \over 2} (\vec{v}_{\rm flow \div \oplus}(t))^2]
= m_a [c^2 + {1 \over 2} (\vec{v}_{\rm flow \div \odot}(t) - 
\vec{v}_{\oplus \div \odot}(t))^2]~~\ , 
\label{annual}
\eeq
where $\bar{\omega}$ is the peak frequency averaged to remove the diurnal 
modulation. $\vec{v}_{\oplus \div \odot}(t)$ follows from the Earth's 
orbital motion.  Eq. (\ref{annual}) then gives $\bar{\omega}(t)$
in terms of $\vec{v}_{\rm flow \div \odot}$, the flow velocity relative 
to the Sun. However, galactic halo models give the velocity
$\vec{v}_\infty$ of the flow long before it reaches the Sun.  
For example, for the 'big flow' of Eq. (\ref{bigflow}), 
$\vec{v}_\infty = \vec{v}_5 - \vec{v}_\odot$ where $\vec{v}_\odot$ 
is the velocity of the Sun relative to the Galaxy.  
$\vec{v}_{\rm flow \div \odot}$ differs from $\vec{v}_\infty$ because 
the flow is affected by the gravitational field of the Sun.  The 
relationship between $\vec{v}_\infty$ and $\vec{v}_{\rm flow \div \odot}$ 
depends on position and hence, as the Earth orbits the Sun, on time. 

One of the main goals of this paper is to derive this relationship.  Note
that a single flow of velocity $\vec{v}_\infty$ far from the Sun, produces
more than one flow in the vicinity of the Sun.  We call $n(\vec{r})$ the
number of flows at position $\vec{r}$.  We want to calculate
$\vec{v}_{{\rm flow} \div \odot,~j}(\vec{r})$ for each of the flows $j = 1
... n(\vec{r})$.  We also want to calculate the density $d_j(\vec{r})$
of each of the flows.

The simplest case is when the Sun is treated as a point mass.  Then, $n=2$
everywhere.  Indeed there are two different ways in which particles coming
with velocity $\vec{v}_\infty$ from far away can reach an arbitrary
location $\vec{r}$.  See Fig. 1.  We will find a caustic line downstream 
of the Sun, on the positive $z$-axis if $\hat{z}$ is the unit vector 
in the direction of $\vec{v}_\infty$.  Indeed, every trajectory passes by
some point on the positive $z$-axis; the latter is a focal line.  We will 
see that the density diverges there as ${1 \over \rho}$ where $\rho$ is
the distance to the $z$-axis.  We call this caustic the 'spike'.

In reality the Sun has finite radius.  This implies that there is a
maximum scattering angle $\Theta_{\rm max}(v_\infty) < 180^\circ$.  As a
result, at large distance $r$ from the Sun, $n=1$ for $\theta >
\Theta_{\rm max}$ and $n=3$ for $\theta < \Theta_{\rm max}$, where
$\theta$ is the polar angle relative to the direction of $\vec{v}_\infty =
\hat{z} v_\infty$.  On the conical surface which separates the $n=1$ and
$n=3$ regions, there is a caustic, which we call the 'skirt'.  The spike
caustic along the $z$-axis downstream from the Sun remains as in the point
mass case.  As the Earth moves around the Sun, it may come close to the
spikes associated with some of the flows and it will go through many of
their skirts.  When it approaches these caustics, the density becomes
larger.  We will see that the profile of the skirt caustic is a functional
of the mass distribution inside the Sun.  We may optimistically envisage a
time where the axion dark matter detector on Earth, monitoring the peak
sizes associated with various flows, provides information about the solar
mass distribution.  

WIMP detectors measure the nuclear recoil energy when a WIMP scatters
elastically off a nucleus in the detector \cite{WIMP,rev}.  The events
caused by a single flow of WIMPs produce a plateau in the distribution of
measured recoil energies.  The edge of the plateau is at the highest
recoil energy compatible with the kinetic energy of the WIMP.  The highest
recoil energy occurs when the collision is back-to-back in the center of
mass.  It is therefore a simple function of the mass of the nucleus, the
mass of the WIMP and the speed of the WIMP in the laboratory frame.  In
principle, a WIMP detector can measure the speed of each of the flows by
measuring the edge of the corresponding plateau in the recoil energy
distribution.  If the detector has directionality \cite{cop}, it may
determine the directions of the flows as well as their speeds. 

All WIMP detectors are presently sensitive to the local dark matter
velocity distribution through the annual modulation effect \cite{ann}.  
The event rate in a WIMP detector is proportional to the flux of WIMPs
through the detector.  (In contrast, the axion to photon conversion rate
in the cavity detector of dark matter axions is proportional to the
density of axions, not their flux.)  The WIMP flux changes in the course
of the year as the velocity of the Earth's orbital motion around the Sun
adds to or subtracts from the velocity of the Sun with respect to the
galactic halo.  

Interestingly, the annual modulation effect in WIMP detectors
distinguishes between the isothermal and caustic ring models of the
Galaxy.  In the isothermal model, the halo has zero average velocity
everywhere.  It has no net angular momentum, no rotation.  The Sun moves
with velocity of order 220 km/s in the direction $\hat{\phi}_G$ of the
galactic disk rotation.  The Earth moves with speed 30 km/s around the Sun
in a plane, called the ecliptic, inclined at approximately $60^\circ$
relative to $\hat{\phi}_G$.  The direction of $\vec{v}_{\oplus \div
\odot}$ is closest to $\hat{\phi}_G$ on about June 2.  The WIMP flux, and
hence the signal in a WIMP detector, is highest then.  It is lowest six
months later, near December 2.  The amplitude of the modulation is of
order $\cos(60^\circ){30~{\rm km/s} \over 220~{\rm km/s}} \simeq 7\%$.  
In the caustic ring model, on the other hand, the halo does have net
angular momentum.  The locally dominant flows, such as
Eq.~(\ref{bigflow}), have velocity components in the $\hat{\phi}_G$
direction of order 470 km/s, more than twice the velocity of the Sun in
that direction.  Relative to the Sun, the average flow of dark matter is
in the $+\hat{\phi}_G$ direction in the caustic ring model, whereas it is
in the $-\hat{\phi}_G$ direction in the isothermal model.  So the annual
modulation in the caustic ring model \cite{bux,ver,gre,gel} has opposite
sign to that of the isothermal model:  the event rate in WIMP detectors is
largest near December 2, lowest near June 2.  Non-isothermal galactic halo
models other than the caustic ring model are discussed in ref.~\cite{ull}.

The DAMA experiment in the Gran Sasso laboratory claims to have observed
the annual modulation of the WIMP signal \cite{dama}.  The DAMA results
are consistent with the prediction of the isothermal model, and
inconsistent with the caustic ring model.  However, almost all of the
region of WIMP mass and elastic scattering cross-section space implied by
the DAMA signal is ruled out by the null result of the CDMS dark matter
search \cite{cdms}.  Some of the region is also ruled out by the null
result of the EDELWEISS experiment \cite{edel}.

At any rate, it is clear that the annual modulation is an important way to
distinguish signal from background in WIMP searches, and that it deserves
the best possible theoretical treatment.  Most calculations and
discussions neglect the effect of the Sun's gravity.  However, as 
we mentioned already, the Sun's gravity effect was analyzed by K. Griest
for the case of the WIMP signal in the isothermal model \cite{gri}.  In
that model the flow of dark matter relative to the Sun produces a wake in
the $-\hat{\phi}_G$ direction.  We may think of it as a spike caustic
except that it is spread out in space because of the large velocity
dispersion ($\sqrt{\vec{v}^2} \simeq 270$ km/s) of the flow.  We 
reanalyze the effect of the Sun's gravity on the annual modulation 
of the WIMP signal in section V.  Our results there are in qualitative
agreement with those of Griest.

In summary, our purpose is to study the effect of Sun's gravity on a flow
of collisionless dark matter in the limit of zero velocity dispersion.  
The effect of finite velocity dispersion can be taken account by
integrating over a distribution of initial velocities.  The results can 
be applied to the flows, or streams, predicted by any specific halo model.

This paper is organized as follows.  In section II, we give a general
discussion of cold (i.e., zero velocity dispersion) flows and associated
caustics.  We derive formulas for the density of such flows that will be
useful in subsequent sections.  In section III, we derive the density and
velocity distributions of a stationary cold dark matter flow past a point
mass.  In that case, all the results are analytical and exact.  In section
IV, we derive the density and velocity distributions of a stationary cold
dark matter flow past a spherically symmetric mass distribution like that
of our Sun.  In this case, the trajectories that go through the Sun are
calculated numerically. The skirt caustic phenomenon appears here.  In
section V, we extend our formalism to dark matter velocities with
arbitrary velocity distribution, and apply the results to the isothermal
model as an example.  In section VI, we summarize our findings.

\section{Density formulas}
\label{sec:density}

Here we derive expressions for the physical space density of a
flow with zero velocity dispersion, using the general approach  
of ref. \cite{PS2}. In such a flow each particle carries a 
3-dimensional label $\vec{\alpha} = (\alpha_{1}, \alpha_{2},
\alpha_{3}).$  The flow is entirely defined by giving the positions 
$\vec{x}(\vec{\alpha},t)$ of all particles at all times.  The velocity 
of particle $\vec{\alpha}$ is 
$\vec{v}(\vec{\alpha},t)=\partial \vec{x}(\vec{\alpha},t)/\partial t.$  
In general, the map $\vec{\alpha} \rightarrow \vec{x}$ is one 
to many.  Hence, at each point $\vec{r}$ of space, there is a 
discrete set of flows with velocities $\vec{v}_{j}(\vec{r},t)=\vec{v}
(\vec{\alpha}_{j}(\vec{r},t),t),\,j=1...n(\vec{r},t),$ where 
$\vec{\alpha}_{j}(\vec{r},t)$ are the solutions of
$\vec{r}=\vec{x}(\vec{\alpha},t).$ $n(\vec{r},t)$ is the number of
distinct flows at $(\vec{r},t).$ The total number of particles is 
\beq 
N = \int d^{3}\alpha \frac{d^{3}N}{d\alpha_{1}d\alpha_{2}d\alpha_{3}}
(\vec{\alpha}) = \int d^{3}r \sum_{j=1}^{n}
\frac{d^{3}N}{d\alpha_{1}d\alpha_{2}d\alpha_{3}}
\frac{1}{\left|\det\left(\frac{\partial\vec{x}}{\partial\vec{\alpha}}\right)
\right|}\,\Biggl|_{\vec{\alpha}=\vec{\alpha}_{j}(\vec{r},t)}.
\label{eq:A1} 
\eeq 
The density of particles in physical space is therefore:  
\beq 
d(\vec{r},t)\,=\,\sum_{j=1}^{n}\,
\frac{d^{3}N}{d\alpha_{1}d\alpha_{2}d\alpha_{3}}
(\vec{\alpha}_{j}(\vec{r},t))\,
\frac{1}{\left|D(\vec{\alpha}_{j}(\vec{r},t),t)\right|}~~~, 
\label{eq:A2}
\eeq 
where 
\beq
D(\vec{\alpha},t)\,\equiv\,\det\left(\frac{\partial\vec{x}}
{\partial\vec{\alpha}}\right).  
\label{eq:A3} 
\eeq 
The formula for the density is manifestly $\vec{\alpha} \rightarrow
\vec{\beta}(\vec{\alpha})$ reparametrization invariant.  

The number of flows $n(\vec{r},t)$ can change from point to point.  
Wherever a change occurs, $n$ changes necessarily by two, i.e. two 
flows are added or two flows disappear.  At those places the map 
$\vec{\alpha}\rightarrow\vec{x}$ is singular, $D$ vanishes and 
the density $d$ is infinite.  Such places are called caustics.  
Caustics are generically surfaces because the condition $D = 0$ 
defines a surface.  Only in degenerate cases are caustics lines 
or points.

A convenient generic parametrization of a zero velocity dispersion flow 
is as follows.  Choose a surface $\cal{S}$ that each particle crosses at
least once in the course of its trajectory.  $\cal{S}$ may be
time--dependent.  Let $\vec{\alpha}=(\alpha_{1},\alpha_{2},t_{0})$
where $t_{0}$ is the time the particle crosses $\cal{S}$ for the
first time, and $(\alpha_{1},\alpha_{2})$ labels the crossing
point on $\cal{S}.$ Then
\beq
\frac{d^{3}N}{d\alpha^{3}}\,=\,\frac{d^{3}N}{d\alpha_{1}d\alpha_{2}dt_0}
\,=\,F(\alpha_{1},\alpha_{2},t_{0})\,
\frac{d^{2}\cal{S}}{d\alpha_{1}d\alpha_{2}}(\alpha_{1},\alpha_{2},t_{0})
\label{eq:A4}
\eeq
where $F(\alpha_{1},\alpha_{2},t_{0})$ is the flux of particles
through the surface.  For a stationary flow it is natural to choose 
a time-independent surface $\cal{S},$ and a time-independent
parametrization $(\alpha_{1},\alpha_{2})$ of the points on
$\cal{S}$.  $F$ and $\frac{d^{2}\cal{S}}{d\alpha_{1}d\alpha_{2}}$ 
are $t_{0}$-independent then, and $\vec{x}(\alpha_{1},\alpha_{2},t_{0};t)
=\vec{x}(\alpha_{1},\alpha_{2},t-t_{0}).$  

In the case of a flow which is initially uniform in space, with density
$d_{\infty}(t_0)$ and velocity $\vec{v}_{\infty}=v_{\infty}(t_0)\hat{z}$,
and which is incident upon a spherically symmetric mass distribution, the
flow is axially symmetric about $\hat{z}.$  It is natural to choose for
$\cal{S}$ a plane perpendicular to $\hat{z}$ at large negative $z$ and 
$(\alpha_{1},\alpha_{2})=(b,\varphi)$ where $b$ is the impact parameter
and $\varphi$ is the azimuthal angle about the
$\hat{z}$--axis.  In this case:
\beq
\vec{x}(b,\varphi,t_{0};t)\,=\,(\hat{x}\,\cos\varphi\,
\,+\,\hat{y}\,\sin\varphi)\,\rho(b,t_{0};t)
\,+\,\hat{z}\,z(b,t_{0};t)
\label{eq:A5}
\eeq
provided we adopt the convention that $\rho$ can be negative as
well as positive and restrict $0\leq\varphi\leq\pi.$  We also let
$b$ have either sign.  [In the usual cylindrical coordinates, where
$\rho\geq 0$ and $0\leq\varphi\leq 2\pi,$ Eq.~(\ref{eq:A5}) would
be invalid because $\varphi$ changes by $\pi$ when the 
$z$-axis is crossed.]  From Eq.~(\ref{eq:A5}) follows:
\beq
\det\left(\frac{\partial\vec{x}}{\partial\vec{\alpha}}\right)
\,=\,\rho\,\det\left(\frac{\partial(\rho,z)}{\partial(b,t_{0})}\right)~~\ .
\label{eq:A6}
\eeq
Moreover
\beq
\frac{d^{3}N}{d\alpha^{3}}\,=\,F(t_0)\,\frac{d^{2}\cal{S}}{db d\varphi}
\,=\,d_{\infty}(t_{0})\,v_{\infty}(t_{0})\,|b|~~\ .
\label{eq:A7}
\eeq
Hence, Eq.~(\ref{eq:A2}) becomes
\beq
d(\rho,z)\,=\,\frac{1}{|\rho|}\,\sum_{j=1}^{n}
\frac{|b|\,d_{\infty}(t_{0})\,v_{\infty}(t_{0})}
{\left|\det\left(\frac{\partial(\rho,z)}{\partial(b,t_{0})}\right)\right|}
\,\Biggl|_{b=b_{j}(\rho,z,t),\,t_{0}=t_{0j}(\rho,z,t)}
\label{eq:A8}
\eeq
where $b_{j}(\rho,z,t)$ and $t_{0j}(\rho,z,t)$ are the solutions
of $\rho(b,t_{0};t)=\rho,\, z(b,t_{0};t)=z.$  If the flow is
stationary, $d_{\infty}$ and $v_{\infty}$ are $t_{0}$--independent,
$\rho(b,t_{0};t)= \rho(b,t-t_{0})$, and $\,z(b,t_{0};t)= z(b,t-t_{0}).$
In that case
\beq
d(\rho,z)\,=\,\frac{d_{\infty}\,v_{\infty}}{|\rho|}\,\sum_{j=1}^{n}
\,\frac{|b|}
{\left|\det\left(\frac{\partial(\rho,z)}{\partial(b,t)}\right)\right|}
\,\Biggl|_{b=b_{j}(\rho,z),\,t=t_{j}(\rho,z)}
\label{eq:A9}
\eeq
where the $t_{j}$ and $b_{j}$ are the solutions of
$\rho(b,t-t_{0})=\rho$ and $z(b,t-t_{0})=z.$

\section{Cold flow past a point mass}
\label{sec:pointmass}

\begin{figure}[t]
\begin{center}
\begin{picture} (300,150)(0,0)
\SetWidth{1.0}
\SetScale{1.}
\GCirc(150,50){5}{0}
\DashLine(10,50)(300,50){3}
\LongArrow(250,60)(250,100)
\LongArrow(250,60)(290,60)
\Curve{(2.38,98.45)(57.97,97.68)(96.59,96.52)(124.66,94.78)
(146.86,92.17)(166.88,88.26)(188.04,82.4)(213.86,73.61)
(248.64,60.43)(298.16,40.68)}
\Curve{(2.2,11.25)(61.56,11.96)(100.65,13.07)(127.52,14.82)
(147.7,17.56)(165.34,21.85)(184.08,28.58)(207.78,39.14)
(241.33,55.71)(291.62,81.68)}
\Text(285,68)[r]{$\hat{z}$}
\Text(255,95)[l]{$\hat{\rho}$}
\end{picture}
\end{center}
\caption{Two trajectories past a spherically symmetric mass distribution.  
The incident flow is assumed to be uniform and to have vanishing velocity
dispersion.  If the scatterer is a point mass, the particles with small
impact parameter are scattered by an angle close to $180^\circ$, and
hence the number of flows $n=2$ everywhere.  If the scatterer is 
distributed over a finite radius, there is a maximum scattering angle 
$\Theta_{\rm max} < 180^\circ$.  In that case, $n=1$ upstream of the 
scatterer, and $n=3$ downstream.}
\label{fig:trajectory}
\end{figure}
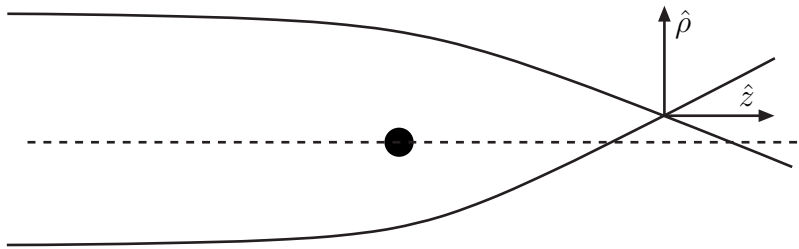

In this section we derive the density and velocity distribution
of collisionless particles in a stationary zero velocity dispersion
flow past a point mass $M$. The flow is axially symmetric as
well as stationary.  Let $d_{\infty}$ and 
$\vec{v}_{\infty}=v_{\infty}\hat{z}$ be the uniform density and 
velocity in the absence of $M$, and $(z,\rho,\varphi)$ cylindrical
coordinates centered on the point mass.  The trajectory of the 
particle with impact parameter $b$ is given in parametric form by
\cite{gol}:
\bea 
t(b,\psi)\, & = &
~\frac{a}{v_{\infty}}~(e~\sinh\psi-\psi)\nonumber \\
z(b,\psi)\, & = & \,a(e~\sinh\psi~+~1~-~\frac{1}{e}\,\exp\psi)\nonumber\\
\rho(b,\psi)~ & = & ~b~(1~-~\frac{1}{e}\,\exp\psi),
\label{eq:2.1}
\eea
where $a\equiv\frac{GM}{v_{\infty}^{2}},$ $\psi$ is the ``eccentric
anomaly'' parameter, and 
$e\equiv\sqrt{1+\left(\frac{b}{a}\right)^{2}}.$ 
Fig.~\ref{fig:trajectory} illustrates a pair of trajectories.  As in 
the previous section, we do not use cylindrical coordinates in the
conventional way.  We allow $\rho$ to have either sign, but restrict 
$\varphi$ to take values between 0 and $\pi.$  Also we let $b$ have 
either sign.  

The velocity components of the flow are:
\bea
v_{\rho}\, = \,\frac{\partial\rho}{\partial\psi}\,
\frac{\partial\psi}{\partial t}\, & = & \,-v_{\infty}\,\frac{b\,\exp\psi}
{e\,a\,(e\,\cosh\psi\,-\,1)}\nonumber \\
v_{z}\, = \,\frac{\partial z}{\partial\psi}\,
\frac{\partial\psi}{\partial t}\, & = & \,v_{\infty}\,
\frac{e\,\cosh\psi\,-\frac{1}{e}\,\exp\psi}{e\,\cosh\psi\,-\,1}~~~\ .
\label{eq:2.2}
\eea
The density distribution is given by Eq. (\ref{eq:A9}).  In the point 
mass case, $n=2$ everywhere.  Upon changing variables $t \rightarrow \psi$, 
we obtain:
\bea
d(\rho,z)\, & = & \,\frac{d_{\infty}\,v_{\infty}}{|\rho|}\,\sum_{j=1}^{2}
\,\frac{|b|\,\left|\frac{dt}{d\psi}\right|}
{\left|\det\left(\frac{\partial(\rho,z)}{\partial(b,\psi)}\right)\right|}
\,\Biggl|_{b=b_{j}(\rho,z),\,\psi=\psi_{j}(\rho,z)}\nonumber\\
& = & \,d_{\infty}\,a\,\sum_{j=1}^{2}
\,\frac{(e\cosh\psi\,-\,1)}{(1\,-\,\frac{1}{e}\exp\psi)\,|D_{2}(b,\psi)|}
\,\Biggl|_{b=b_{j}(\rho,z),\,\psi=\psi_{j}(\rho,z)}
\label{eq:2.4}
\eea
where $b_{j}(\rho,z)$ and $\psi_{j}(\rho,z)$ are the two solutions
of $\rho(b,\psi)=\rho$ and $z(b,\psi)=z$, and
\bea
D_{2}(b,\psi)\,& \equiv & \,\det\left(\frac{\partial(\rho,z)}
{\partial(b,\psi)}\right) \nonumber \\
& = & \frac{a}{2}\left(e\,\exp(-\psi)\,-\,1\,+\,
\left(e\,-\,\frac{2}{e}\right)\,\exp\psi\,+\,\exp(2\psi)\right).
\label{eq:2.5}
\eea
Eqs.~(\ref{eq:2.1}) can be inverted: 
\bea
b_{\pm}(\rho,z)\, & = &
\,\frac{\rho}{2}\left(1\,\pm\,\sqrt{1\,+\,y}\right)\nonumber \\
\psi_{\pm}(\rho,z)\, & = & \ln\left[
e_{\pm}\,\frac{1}{y}\,\left(1\mp\,\sqrt{1\,+\,y}\right)^{2}\right]
\label{eq:2.6}
\eea
where $e_\pm = \sqrt{1 + ({b_\pm \over a})^2}~$ and
\beq
y\,\equiv\,\frac{4a}{\rho^{2}}~(r~+~z)
=\,\frac{2a}{r\,\sin^{2}(\theta/2)}~~~,
\label{eq:2.7}
\eeq
with $r=\sqrt{\rho^{2}+z^{2}}$ and $\theta=\tan^{-1}(\rho/z)$.
Henceforth we label the two flows with $j=\pm.$  Inserting 
Eqs.~(\ref{eq:2.6}) into Eqs.~(\ref{eq:2.5}) and (\ref{eq:2.4}) yields:
\beq
D_{2\pm}\,=\,r\,\frac{2\,\sqrt{1\,+\,y}}{\sqrt{1\,+\,y}\,\pm\,1}
\label{eq:2.8}
\eeq
and
\beq
d_{\pm}\,=\,\frac{d_{\infty}}{4}\,\left(
\sqrt{1\,+\,y}\,+\frac{1}{\sqrt{1\,+\,y}}\,\pm\,2\right).
\label{eq:2.9}
\eeq
Note that $d_{+}\,-\,d_{-}\,=\,d_{\infty}$ everywhere.
The velocities are:
\bea
v_{\rho\pm}\,& = & \,-\,\frac{v_{\infty}}{r}\,\left(
b_{\pm}\,-\,\rho\right)\nonumber \\
v_{z\pm}\,& = & \,v_{\infty}\,\left(1\,+\,
\frac{a\,\rho}{r\,b_{\pm}}\right)~~~~~\ .
\label{eq:2.10}
\eea
The total density
\beq
d~=~d_+~+~d_-~=\,\frac{d_{\infty}}{2}\,\left(
\sqrt{1+\frac{2\,a}{r\,\sin^{2}(\theta/2)}}
\,+\,{1 \over \sqrt{1+\frac{2\,a}{r\,\sin^{2}(\theta/2)}}}\right)~~\ .
\label{eq:2.11}
\eeq
It diverges on the positive $\hat{z}$--axis as 
\beq
d\,\simeq\,d_{\infty}\,\frac{\sqrt{2\,a\,z}}{|\rho|}~~~\ .
\label{eq:2.12}
\eeq
This divergence occurs because the flow is focussed onto the positive
$\hat{z}$--axis by the point mass' gravity.  Each trajectory crosses 
the positive $\hat{z}$--axis, at $z = {b^2 \over 2a}$. The 
positive $\hat{z}$--axis is the location of a line caustic, which we 
call the 'spike'.  Eq.~(\ref{eq:2.12}) is valid also for a finite size 
but spherically symmetric mass, at $z$ large enough that the 
trajectories there did not pass through the mass.  

As we remarked already in section II, caustics are generally surfaces
rather than lines or points \cite{PS2}.  A line or point caustic is a
degenerate case.  The caustic of Eq.~(\ref{eq:2.12}) is a line only
because of the axial symmetry of the flow.  If less symmetry is present -
e.g. if the mass has a quadrupole moment, or if the initial velocity field
of the flow is inhomogeneous - there is still a caustic downstream of the
mass but it is a surface.  Since it can collapse to a line, the surface is
a tube of some sort.  The structure of the tube is a problem left for
future investigations.

\section{Cold flow past a spherically symmetric mass}
\label{sec:solar}

In this section we derive the density and 
velocity distribution
of a zero velocity dispersion flow past a spherically 
symmetric mass distribution of total mass $M_{\odot}.$  
The mass distribution is characterized by a radius $R_{\odot}$
outside of which the density rapidly approaches zero.
The trajectories 
passing through the mass must be calculated numerically.  
As before, the only force acting is the gravity of $M_{\odot}.$
The density $d_{\infty}$ 
and velocity $\vec{v}_{\infty}=v_{\infty}\hat{z}$ 
of the flow long before encountering the mass
distribution are assumed to be constant in both space and time.
Axial symmetry is preserved and the flow is stationary. 

The density and velocity distributions derived 
in the previous section (\ref{sec:pointmass}) are valid here
provided the trajectories did not pass through the mass
distribution at any point in their
past.  Defining $b_{\odot}$ to be the impact 
parameter of the trajectory grazing the surface of 
the mass distribution, all results of the point mass case 
hold for trajectories labeled by $b>b_{\odot}.$  The $b>b_{\odot}$
trajectories are given in parametric form by 
Eqs.~(\ref{eq:2.1}) which implies $r=a(e\cosh\psi-1).$  
The point of closest approach to the mass 
is where $\psi=0.$  
Setting $\psi=0$ and $r=R_{\odot}$ gives
the grazing impact parameter
\beq
b_{\odot}=R_{\odot}\sqrt{1+2a/R_{\odot}}\;.
\label{eq:grazing}
\eeq
On the other hand, all trajectories for $b<b_{\odot}$ 
will pass through the mass distribution at some point 
in their motion.  

Here it is convenient to work in spherical
coordinates $(r,\theta,\varphi).$ 
As in the previous section the azimuth $\varphi$
is restricted to be between $0$ and $\pi.$  We
measure the polar angle $\theta$ from the $+\hat{z}$--direction
and it takes values between $-\pi$ and $+\pi.$  In this
system of coordinates the polar
angle smoothly changes sign as a trajectory crosses the symmetry
axis downstream from the mass.  As before, we let $b$ have
either sign.

We derive the density distribution of 
a stationary flow with axial symmetry in spherical
coordinates.  Analogously to Eqs.~(\ref{eq:A6}) and (\ref{eq:A9}), we have 
\beq
\det\left(\frac{\partial\vec{x}}{\partial\vec{\alpha}}\right)
=r^2\sin\theta\det\left(\frac{\partial(r,\theta)}{\partial(b,t_{0})}\right)
\eeq
and
\beq
d(r,\theta)\,=\,\frac{d_{\infty}\,v_{\infty}}{r^2|\sin\theta|}
\,\sum_{j=1}^{n}\,\frac{|b|}
{\left|\det\left(\frac{\partial(r,\theta)}{\partial(b,t)}\right)\right|}
\,\Biggl|_{b=b_{j}(r,\theta),\,t=t_{j}(r,\theta)}~.
\label{eq:3.2}
\eeq
The index $j$ counts the flows.  In the point mass case,
particles with impact parameter close to zero are scattered
by an angle close to $180^{\circ}.$  In that case, $n=2$ everywhere.
On the other hand, if the mass has finite size, there is
a maximum scattering angle $|\Theta_{\rm{max}}|<180^{\circ}.$
This angle defines a cone.  Far from the mass, $n=1$ for 
$|\theta|>|\Theta_{\rm{max}}|$ whereas $n=3$ for
$|\theta|<|\Theta_{\rm{max}}|.$  Since the number of flows changes
on the surface of the cone, we expect a caustic there.

We break the sum over $j$ into two parts. The first sum
is over the flows for which we have calculated an exact result:
the $b>b_{\odot}$ flows at any time in their motion and 
the $b<b_{\odot}$ flows before entering the mass distribution.  
We label these flows by the index $k~(=1,...,n_k).$ 
The second sum is over the flows which must be calculated
numerically, namely the $b<b_{\odot}$ flows at any time
after they have entered the mass distribution.  We label
these flows by the index $l~(=1,...,n_l).$  The density is then    
\beq
d(r,\theta)\,=\,\sum_{k=1}^{n_k}d_k(r,\theta)
\,+\,\frac{d_{\infty}v_{\infty}}{r^2|\sin\theta|}\,\sum_{l=1}^{n_l} \,\frac{|b|}
{\left|\det\left(\frac{\partial(r,\theta)}{\partial(b,t)}\right)\right|}
\,\Biggl|_{b=b_{l}(r,\theta),\,t=t_{l}(r,\theta)}
\label{eq:3.3}
\eeq
where the $d_k(r,\theta)$ expressions were derived in 
section \ref{sec:pointmass} and the terms in the sum
over $l$ we address next.  Although we have not shown
it explicitly Eq.~(\ref{eq:3.3}), $n_k$ and $n_l$ are dependent
upon $(r,\theta).$  We will see that  
$(n_k,n_l)=(1,0),(1,2),(2,1),(0,1)$ or $(0,3),$ depending on position.

\begin{figure}
\epsfxsize=5in
\centerline{\epsfbox{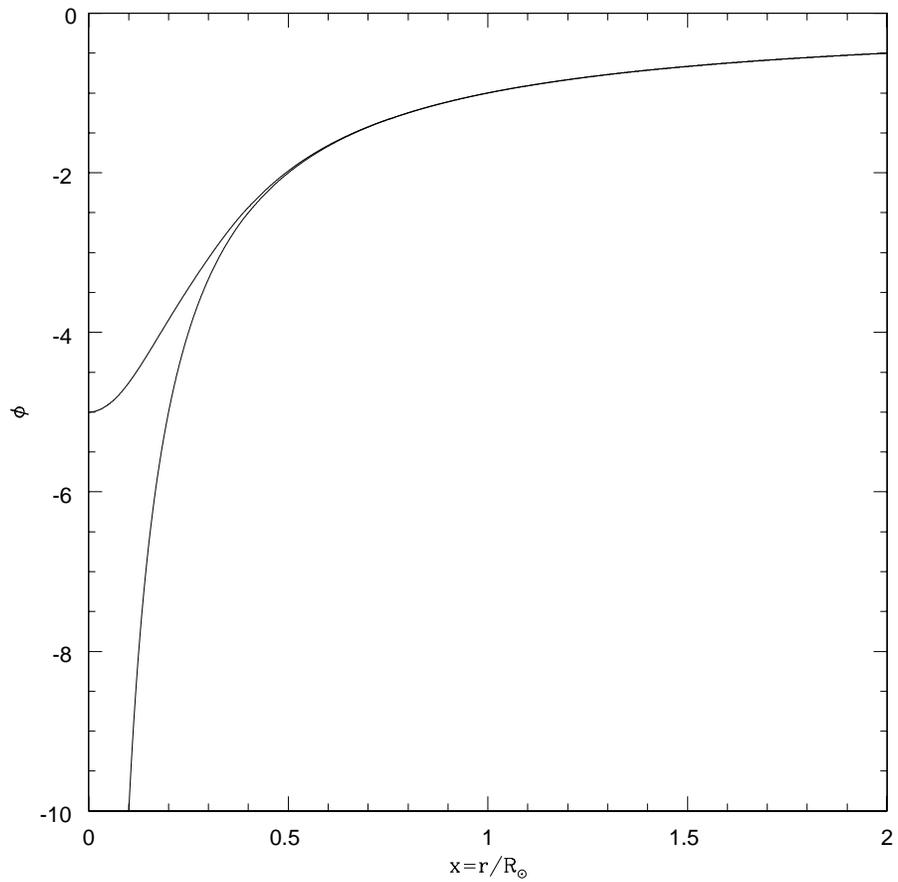}}
\caption{A plot of the dimensionless Solar potential $\phi(x)$ 
given in Eq.~(\ref{eq:potential}) and compared with
the point mass potential $\phi(x)=-1/x$.}
\label{fig:solarpotential}
\end{figure}

To simplify the determinant, while
exposing the caustic features intrinsic to
Eq.~(\ref{eq:3.3}), we take the limit where
$r>>R_{\odot}, a.$   In that case, for the $l$--flows   
\beq
\det\left(\frac{\partial(r,\theta)}{\partial(b,t)}\right)
\,\simeq\; v_{\infty}\frac{\partial\theta}{\partial b}~~.
\label{eq:3.4}
\eeq
Hence
\beq
d_l(r,\theta)\,\simeq\,\frac{d_{\infty}}{r^2|\sin\theta|}\,
\,\frac{|b_l|}{\left|\frac{d\theta}{db_l}\right|}
\label{eq:density_l}
\eeq
and the numerical computation of the second term in
Eq.~(\ref{eq:3.3}) becomes straightforward.
For the $l$--flows at $r>>R_{\odot}$ and $a,$ $~|\theta (b)|$ 
is the scattering angle
for the trajectory of impact parameter $|b|.$
The scattering angle of a particle incident at
impact parameter $b$ with initial kinetic energy $E$ upon a
central potential $V(u)$ is \cite{gol}
\beq
\Theta(|b|)\,=\,~\pi\,-\,2\,\int_{0}^{u_{\rm{max}}}
\,\frac{|b|\,du}{\sqrt{1\,-\,V(u)/E\,-\,b^{2}u^{2}}},
\label{eq:theta}
\eeq
where $u\equiv 1/r$ and $u_{\rm{max}}= 1/r_{\rm{min}}.$ 
We express the potential as $V(x)=GM_{\odot}\phi(x)/R_{\odot}$
with $x\equiv r/R_{\odot}.$
We choose the potential of the mass distribution to be
\beq
\phi(x) = -\left[\frac{\alpha+\beta x^{6}}{1+\gamma x^{2}
+\beta x^{7}}\right],
\label{eq:potential}
\eeq  
setting $\alpha=5,~\beta=5000,$ and $\gamma=8$ to give a close
fit to the solar mass distribution \cite{Sun}.
Thus, all of the following calculations should be considered
relevant to the Sun.  
In Eq.~(\ref{eq:potential}), $\phi\simeq-1/x$ for $x>1,$ 
the point mass limit, and $\phi\simeq -\alpha+\gamma x^2$ for $x<<1$ 
reflecting the fact that the density in the solar core is constant. 
The potential (\ref{eq:potential}) is 
plotted in Fig.~\ref{fig:solarpotential} in comparison
to $\phi=-1/x.$

\begin{figure}
\epsfxsize=5in
\centerline{\epsfbox{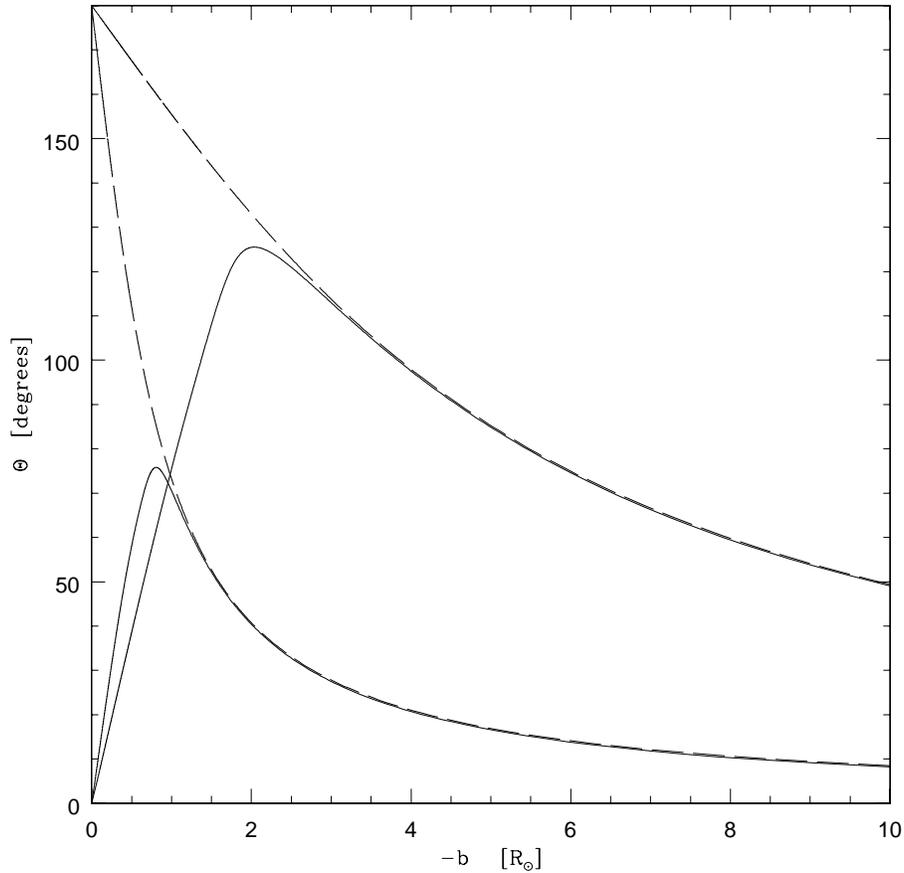}}
\caption{The scattering angle for dark matter incident upon the 
Sun as a function of impact parameter, for flows of initial velocity
$200$ km/sec and $500$ km/sec.  The solid curves are for the distributed
mass described in section IV and the dashed curves are for the point
mass.  The flows with lower initial velocities have larger 
scattering angle at large impact parameter.}
\label{fig:dmscatter}
\end{figure}

We calculate the scattering angle $\Theta(|b|)$ for the potential 
of Eq.~(\ref{eq:potential}). 
For comparison, the scattering angle for the $\phi=-1/x$ 
potential is the exact result
\beq
\Theta(b)~=~-~2~\arcsin\left(\frac{1}{e}\right)~{\rm{sign}}(b)\;,
\label{eq:scat_exact}
\eeq 
in which $\Theta(b)$ can take values between 
$-\pi$ and $+\pi.$ 
We plot $\Theta(b)$ for the distributed mass and point 
mass cases in Fig.~\ref{fig:dmscatter} 
calculated for two flow velocities,
$v_{\infty}=200$ km/sec and $500$ km/sec.  

In Fig.~\ref{fig:dmscatter} there 
is a maximum scattering angle $\Theta_{\rm{max}}~<~\pi.$ 
For $0~<~\theta~<~\Theta_{\rm{max}},$ there are
two flows labeled by different values of $b$
with ${\rm{sign}}(b)=-\rm{sign}(\theta).$
In addition, there is a flow for which $b$ and
$\theta$ have the same sign.  For large $r,$ the third 
flow is in the sum over $k.$  In the notation of 
section \ref{sec:pointmass}, it has impact parameter $b_{+}.$
So $n=3,$ for large $r$ and
$|\theta|<\Theta_{\rm{max}}.$  For 
$\Theta_{\rm{max}}<|\theta|<\pi,$ the two flows of 
Fig.~\ref{fig:dmscatter} are absent but the $b_{+}$
flow is still present, so $n=1$ there.  

\begin{figure}
\epsfxsize=5in
\centerline{\epsfbox{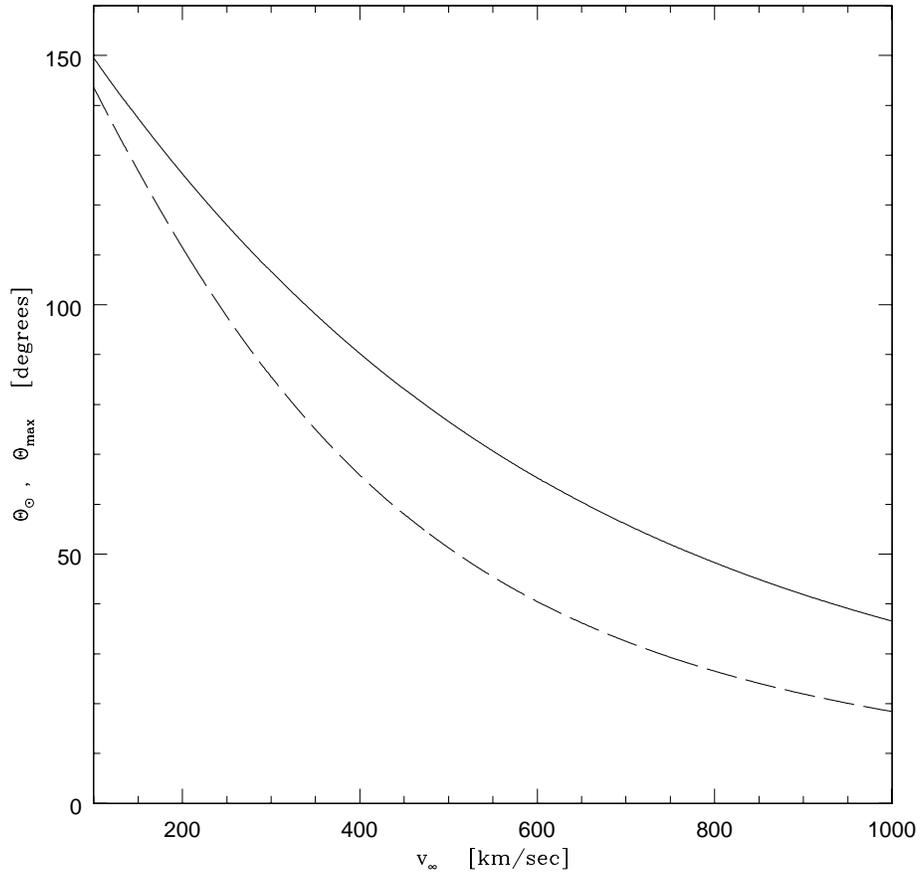}}
\caption{The maximum scattering angle $\Theta_{\rm{max}}$ (solid line)
and the scattering angle $\Theta_\odot$ of the grazing trajectory 
(dashed line) as a function of initial flow velocity $v_{\infty}$, 
for the solar model described in section IV.}
\label{fig:dmthmax}
\end{figure}

Near the cone $|\theta|=\Theta_{\rm{max}}$ is a caustic
surface which we call the ``skirt.''
The determinant in Eq.~(\ref{eq:3.3}) vanishes there
since $d\theta/db=0.$
We calculated the maximum scattering angle $\Theta_{\rm{max}}$
as a function of flow velocity $v_{\infty}$  
ranging from $100$ km/sec to $1000$ km/sec.
See Fig.~\ref{fig:dmthmax} for the result.

\begin{figure}
\epsfxsize=5in
\centerline{\epsfbox{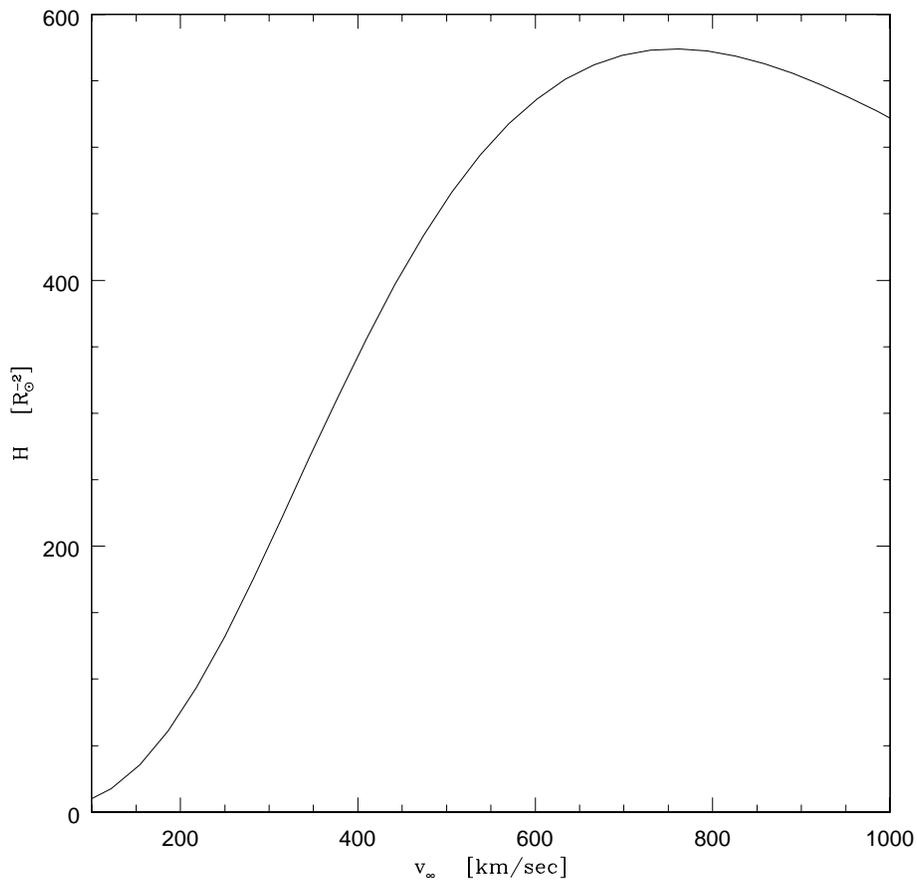}}
\caption{The parameter $H\equiv -\frac{d^2\Theta}{db^2}|_{b=b_c}$ 
as a function of the initial flow velocity $v_{\infty}$, for the 
solar model described in section IV.  The strength of the skirt 
caustic is proportional to $H^{-1/2}.$}
\label{fig:dmd2thdb2}
\end{figure}

\begin{figure}
\epsfxsize=5in
\centerline{\epsfbox{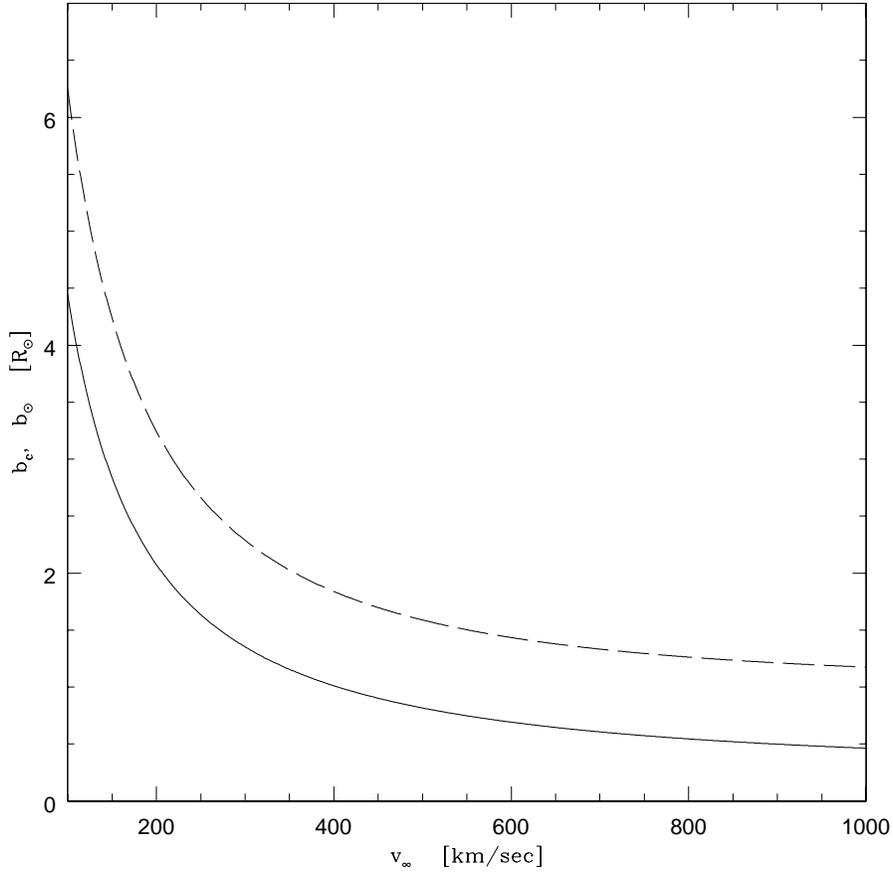}}
\caption{The impact parameter $b_c$ of the trajectory of maximum
scattering angle (solid line) and the impact parameter $b_\odot$
of the grazing trajectory (dashed line) as a function of initial 
flow velocity $v_{\infty}$, for the solar model described in section
IV.  The strength of the skirt caustic is proportional to $b_c$.}
\label{fig:dmbcrit}
\end{figure}

We now examine the density Eq.~(\ref{eq:3.3}) 
near the skirt caustic for angles $\theta<\Theta_{\rm{max}}.$
The function $\Theta(b)$
can be approximated near $\Theta_{\rm{max}}$ as
\beq
\Theta(b)\simeq \Theta_{\rm{max}}-\frac{1}{2}H(b-b_c)^2
\label{eq:hoss}
\eeq
where $b_c$ is the impact parameter
for the maximum scattering angle, and
$H\equiv-\frac{d^2\Theta}{db^2}|_{b=b_c}.$
For $\theta$ near to but smaller than $\Theta_{\rm{max}},$ 
we have $n_k=1$ and $n_l=2$ in Eq.~(\ref{eq:3.3}).
The $k=1$ flow is the one labeled $+$ in section \ref{sec:pointmass}.  
Using Eqs.~(\ref{eq:3.4}) and (\ref{eq:hoss}),
we get the approximate density near the skirt
\bea
d_{\rm{skirt}} & \simeq~d_{+}~+~\frac{d_{\infty}~b_c}
{r^2\sin\Theta_{\rm{max}}}
\sqrt{\frac{2}{H~(\Theta_{\rm{max}}-\theta)}} 
& ~~~ \rm{for}~~ \theta<\Theta_{\rm{max}}\\
 & =~d_{+}~~~~~~~~~~~~~~~~~~~~~~~~~~~~~~~~~
& ~~~ \rm{for}~~ \theta>\Theta_{\rm{max}}\;.
\label{eq:skirt}
\eea
A generic feature of 
caustic surfaces is evident in Eq.~(\ref{eq:skirt}), namely that
the density diverges near the surface
as $1/\sqrt{\Theta_{\rm{max}}-\theta}$ 
when the skirt is approached from $\theta~<~\Theta_{\rm{max}}$
side.   The density remains constant (equal
to $d_{+}$) when the surface is
approached from $\theta~>~\Theta_{\rm{max}}.$
The strength of the caustic is proportional
to $b_c/r^2\sin\Theta_{\rm{max}}\sqrt{H}.$
We plot $H$ and $b_c$ over a range of flow velocities
in Figs.~\ref{fig:dmd2thdb2} and \ref{fig:dmbcrit},
respectively.  For comparison, we plot $b_c$ alongside
the grazing impact parameter $b_{\odot}.$ 

Moving away from the skirt toward smaller values of
$|\theta|,$ an angle $\Theta_{\odot}$
will be reached where the flows change from $n_k=1,$ $n_l=2$
to $n_k=2,$ $n_l=1.$ 
$\Theta_{\odot}=\Theta(b_{\odot})$ is the scattering
angle of the grazing trajectories.  Using Eq.~(\ref{eq:grazing})
gives
\beq
\Theta_{\odot}~=~-~2~\arcsin\left(\frac{a}{a+R_{\odot}}\right)
~{\rm{sign}}(b_{\odot})\;.
\label{eq:thetasolar}
\eeq
For $|\theta|<|\Theta_{\odot}|,$ only one flow passed 
through the solar interior.  We plot $\Theta_{\odot}$
in Fig.~\ref{fig:dmthmax} as a function of the flow
velocity $v_{\infty}.$

For $r>>R_{\odot}$ and $a,$
\beq
\ba{rl}
(n_k,n_l)~=~(1,0)~~&\rm{for}~|\theta|>|\Theta_{\rm{max}}|\\
=~(1,2)~~&\rm{for}~|\Theta_{\odot}|<|\theta|<|\Theta_{\rm{max}}|\\
=~(2,1)~~&\rm{for}~|\theta|<|\Theta_{\odot}|~~. \ea  
\eeq
There are two regions in or near the Sun where different
$(n_k,n_l)$ values arise.
First, $(n_k,n_l)=(0,1)$ on the interior
of the Sun on the upstream side.
Second, there is a region near the Sun on the downstream side 
where $(n_k,n_l)=(0,3).$  The flow properties in these regions 
are calculable with Eq.~(\ref{eq:3.3}).   We leave that
task for future investigations.

The velocity distributions for the $k$--flows 
are given in Eq.~(\ref{eq:2.10}).
For the $l$--flows, using energy and angular momentum conservation,
\bea
v_{r_l}~&=&~v_{\infty}~\sqrt{1+\frac{2a}{r}
-\left(\frac{b_l}{r}\right)^{2}}~~,\nonumber \\
v_{\theta_l}~&=&~v_{\infty}\frac{b_l}{r}~~.
\label{eq:l-velocities}
\eea
For the $l$--flows at large $r,$ $\theta$ is the scattering
angle $\Theta.$  Hence, the $b_l(\theta)$ are 
given by Fig.~\ref{fig:dmscatter}.

\begin{figure}
\epsfxsize=5in
\centerline{\epsfbox{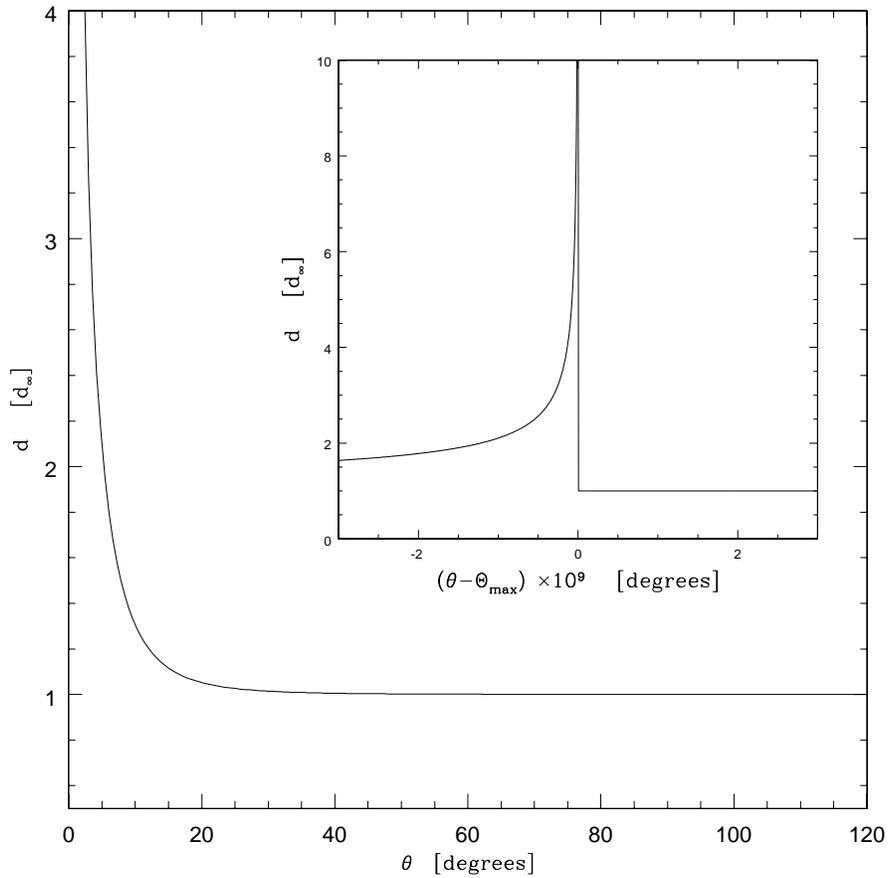}}
\caption{The density of a cold collisionless flow past the Sun
as a function of polar angle for $r=1$~AU and $v_\infty$ = 255 km/s.
The divergence at $\theta = 0$ is the spike caustic.  The skirt 
caustic is at $\theta = 115^\circ$.  It is invisible on the scale 
of the main plot but is shown in the inset.}
\label{fig:dmdensity}
\end{figure}

Using $\Theta(b)$ it is straight--forward to calculate
the full expression for the density distribution at
large $r$ including the flows through the Sun.
This result is given in Fig.~\ref{fig:dmdensity} for
the mean Earth--Sun distance.  The density field of a $b<b_{\odot}$ flow, 
at large $r,$ is smaller than the sum of the 
$b>b_{\odot}$ flows, roughly by the factor
$\frac{b_{\odot}^{2}}{r^{2}}.$   Away from the caustic skirt,
the flows through the Sun contribute only $4\times 10^{-3}$
to $4\times 10^{-5}$
of the total density at the Earth's location for velocities 
ranging from $v_{\infty}=100$ km/sec to 1000 km/sec.

The limit $R_{\odot}\rightarrow 0$ seems paradoxical.
In section \ref{sec:pointmass} we found that
$n=2$ everywhere, whereas here $n=1$
for $|\theta|>|\Theta_{\rm{max}}|$ and $n=3$ for 
$|\theta|<|\Theta_{\rm{max}}|.$  What happens is that when
$R_{\odot}\rightarrow 0$ and $\Theta_{\rm{max}}\rightarrow\pi$ 
one of the three flows acquires vanishing density.  The
flow that disappears is the one with the smallest impact 
parameter for given scattering angle in Fig.~\ref{fig:dmscatter}.
The flows that survive are the flow with the largest
impact parameter for given scattering angle in 
Fig.~\ref{fig:dmscatter} and the flow with 
sign($\theta$)=sign($b$), i.e., the flow labeled $+$ in
section \ref{sec:pointmass}.

To conclude this section, we briefly discuss
the relationship between the density distribution
formulas for a flow past a point mass and the Rutherford 
differential cross--section formula.
The differential scattering cross--section is 
\beq
\frac{d\sigma}{d\Omega}(\theta)=\frac{b}{\sin\theta}
\left|\frac{db}{d\theta}\right|\;,
\label{eq:B1}
\eeq
and the Rutherford formula for scattering 
in a $r^{-1}$ potential is
\beq
\frac{d\sigma}{d\Omega}(\theta)_{\rm{R}}
=\frac{a^2}{4\sin^4(\theta/2)},
\label{eq:B2}
\eeq
where $a\equiv GM/v_{\infty}^{2}$ as before.
Consider only the ``$-$ flow,'' 
in the notation of section \ref{sec:pointmass},
in the limit where $r>>a$ and $\theta\neq 0.$
As shown above, the determinant 
then simplifies, 
$\left|\det\left(\frac{\partial(r,\theta)}{\partial(b,t)}\right)\right|
\rightarrow v_{\infty}\left|\frac{d\theta}{db}\right|,$
and the density can be written in terms of
the Rutherford formula Eq.~(\ref{eq:B2}): 
\beq
d_{-}(r,\theta)\,=\,\frac{d_{\infty}}{r^{2}}\,
\frac{b}{\sin\theta}\left|\frac{db}{d\theta}\right|
=\,\frac{d_{\infty}}{r^{2}}\,
\frac{d\sigma}{d\Omega}(\theta)_{\rm{R}}\;.
\label{eq:B5}
\eeq
Using a fact derived in section \ref{sec:pointmass}, namely that
$d_{+}-d_{-}=d_{\infty}$ everywhere, we get
\beq
d_{+}(r,\theta)\,=\,d_{\infty}\left(1+
\frac{a^2}{4 r^{2} \sin^4(\theta/2)}\right).
\label{eq:B6}
\eeq
One can verify that the Eqs.~(\ref{eq:B5}) and (\ref{eq:B6})
agree with the exact results for $d_{\pm}$ derived in 
section \ref{sec:pointmass}.  Taking the $r\rightarrow\infty$
limit of Eq.~(\ref{eq:2.9}) at fixed $\theta\neq 0$ and keeping
the terms to leading order yields 
Eqs.~(\ref{eq:B5}) and (\ref{eq:B6}).

\section{Hot flow past a point mass}
\label{sec:boostedisotherm}

In many discussions of dark matter detection on Earth, it is assumed 
that the distribution in the galactic halo is isothermal.  The local 
velocity distribution is then
\beq
f(\vec{v}_\infty) = \frac{d_\infty}{(\sqrt{\pi}\sigma)^{3}}
\exp[-\frac{1}{\sigma^{2}}(\vec{v}_\infty+\vec{v}_{\odot})^{2}]
\label{eq:5.1}
\eeq
in the absence of the Sun, for an observer moving with the
Sun's velocity $\vec{v}_{\odot}$.  
$\sigma=\sqrt{\frac{2}{3}<v^{2}>}\simeq~220$ km/s, where $<v^{2}>^{1/2}$ 
is the velocity dispersion of the galactic halo.  $\sigma$ and 
$v_{\odot}$ are approximately equal. As was mentioned in 
the Introduction, there is little theoretical motivation for 
assuming an isothermal distribution.  Indeed most of our paper 
is concerned with discrete flows, which we believe to be the  
relevant description.  However our methods can also be applied 
to smooth velocity distributions such as Eq.~(\ref{eq:5.1}).  This 
is done in this section.  We neglect here the Sun's size, treating
it as a point mass.

If the velocity distribution in the absence of the Sun is
$f(\vec{v}_\infty)$ independently of position, what is the 
position-dependent velocity distribution $f_{\odot}(\vec{r},\vec{v})$ 
when the effect of the Sun's gravity is included?  The answer to this 
seemingly daunting question is \cite{dan,gri}:
\beq
f_{\odot}(\vec{r},\vec{v})=f(\vec{v}_\infty(\vec{r},\vec{v}))
\label{eq:5.2}
\eeq
where $\vec{v}_\infty(\vec{r},\vec{v})$ is the initial velocity
of the trajectory that passes by position $\vec{r}$ with
velocity $\vec{v}.$  Indeed, Liouville's theorem states
that phase-space density is conserved along particle 
trajectories.  Since $f(\vec{v}_\infty)$ is the initial 
phase-space density, when the particles are far from
the Sun, and $f_{\odot}(\vec{r},\vec{v})$ is the final 
phase-space density, Eq.~(\ref{eq:5.2}) follows.

To obtain $\vec{v}_\infty(\vec{r},\vec{v}),$ we exploit the
constants of the motion:
\bea
E & = & \frac{1}{2}v^{2} - \frac{G M_{\odot}}{r}\nonumber \\
\vec{l} & = & \vec{r} \times \vec{v} \nonumber \\ 
\vec{A} & = & \vec{v} \times (\vec{r} \times \vec{v}) 
- G M_{\odot} \hat{r}~~\ . 
\label{eq:5.3}
\eea
$E,~\vec{l},$ and $\vec{A}$ are the energy, angular
momentum and Laplace--Runge--Lenz vector, per unit
mass in each case.  In the initial state,
\bea
E & = & \frac{1}{2}v_\infty^{2} \nonumber \\
\vec{l} & = & - b v_\infty \hat{\varphi} \nonumber \\ 
\vec{A} & = & b v_\infty^{2} \hat{\rho} + G M_{\odot} \hat{z}. 
\label{eq:5.4}
\eea
where $b$ is the impact parameter, $\hat{z}$ is in the direction
of the initial velocity $(\vec{v}_\infty=v_\infty\hat{z}),$  
$\hat{\varphi}$ is perpendicular to the plane of the trajectory
and $\hat{\rho}=\hat{\varphi}\times\hat{z}.$ The notation is 
consistent with that of section III.  Since 
\beq
v_\infty = + \sqrt{v^{2}-\frac{2 G M_{\odot}}{r}},
\label{eq:5.5}
\eeq
it only remains to express $\hat{z}$ in terms of $\vec{r}$ and 
$\vec{v}.$ We have:
\beq
\vec{A}\times\vec{l} = + b v_\infty^3(a \hat{\rho} - b \hat{z}) 
\label{eq:5.6}
\eeq
where $a=\frac{G M_{\odot}}{v_\infty^{2}},$ as before.  Hence
\beq
\hat{z}=\frac{1}{(a^{2}+b^{2})v_\infty^{2}}(a \vec{A} 
-\frac{1}{v_\infty} \vec{A}\times\vec{l})~~\ .
\label{eq:5.7}
\eeq
Substituting Eqs.~(\ref{eq:5.3}) into Eq.~(\ref{eq:5.7}), we obtain
after some algebra:
\beq
\vec{v}_\infty(\vec{r},\vec{v})=\frac{1}{a^{2}v_\infty^{2}+l^{2}}
\left\{ \vec{v} \left[ l^{2}-av_\infty^{2}r-av_\infty(\vec{r}\cdot\vec{v})
\right] + \vec{r}av_\infty^{2} \left[ \frac{1}{r}(\vec{r}\cdot\vec{v})
+\frac{v^{2}}{v_\infty}-\frac{a v_\infty}{r} \right] \right\}
\label{eq:5.8}
\eeq
where
$l^{2}=r^{2}v^{2}-(\vec{r}\cdot\vec{v})^2$ and $v_\infty$ is given in
terms of $\vec{r}$ and $\vec{v}$ by Eq.~(\ref{eq:5.5}).

If the isothermal velocity distribution (\ref{eq:5.1}) is assumed, we 
have therefore:
\beq
f_{\odot}(\vec{r},\vec{v})=\frac{d_{\infty}}{(\sqrt{\pi}\sigma)^{3}}
\exp[-\frac{1}{\sigma^{2}}v_{G}^{2}(\vec{r},\vec{v})]
\Theta(v_\infty^{2}(\vec{r},\vec{v}))
\label{eq:5.9}
\eeq
where $\Theta(x) = 0$ for $x<0$, 1 for $x>0$, and
\bea
v_{G}^{2}(\vec{r}, \vec{v}) & = &
(\vec{v}_{\odot}+\vec{v}_\infty(\vec{r},\vec{v}))^{2} 
=  \vec{v}_{\odot}^{2}+\vec{v}_{\infty}^{2} + 
2v_{\odot}\frac{1}{a^{2}v_\infty^{2}+l^{2}} \cdot\nonumber\\
& \cdot & \left\{v\cos\alpha\left[l^{2}-av_\infty^{2}r-a 
v_\infty(\vec{r}\cdot\vec{v})\right]
+a v_\infty^{2}\cos\beta\left[\vec{r}\cdot\vec{v}
+\frac{v^{2}r}{v_\infty}-a v_\infty\right]\right\}~~~,
\label{eq:5.10}
\eea
where $\alpha$ is the angle between $\vec{v}_{\odot}$ and the dark matter
velocity $\vec{v}$ and $\beta$ is the angle between $\vec{v}_{\odot}$ and
the position $\vec{r}$ of the observer relative to the Sun.  A formula for 
$v_{G}^{2}(\vec{r},\vec{v})$ was obtained by J.M.A. Danby and G.L. Camm
\cite{dan} over 40 years ago.  K. Griest \cite{gri} obtained a formula 
which differs from that of Danby and Camm.  Our formula appears 
different from both that of Danby and Camm, and that of Griest.
We verified that the RHS of Eq.~(\ref{eq:5.8}) is indeed $v_{\infty}^{2}$,
and that Eq.~(\ref{eq:5.8}) checks out in the $l=0$ and 
$l\rightarrow\infty$ limits, in the $a=0$ limit, and when
$\vec{v}_{\infty}\parallel\vec{r}$.

The annual modulation of the event rate in WIMP dark matter searches is
caused for the most part by the seasonal variation of the flux of dark
matter particles as the Earth's orbital motion adds to or subtracts from the
motion of the Sun through the galactic halo.  As was discussed in the
Introduction, there is an additional contribution because the Sun's gravity
modifies the density and velocity distributions in the solar neighborhood.
The combined effects of the Earth's orbital motion and the Sun's gravity
can be computed if we know the distribution of dark matter in the Earth's
frame of reference.  Following Griest, that is obtained by simply
replacing $\vec{v}_{\odot}$ in Eqs.~(\ref{eq:5.9}) and (\ref{eq:5.10})  
by $\vec{v}_{\odot}+\vec{v}_{\oplus}$ where $\vec{v}_{\oplus}$ is the
velocity of the Earth relative to the Sun.  The event rate depends also 
on the WIMP mass and such experimental variables as the detector
threshold, the target mass and, of course, the scattering
cross--section.

\begin{figure}
\epsfxsize=5in
\centerline{\epsfbox{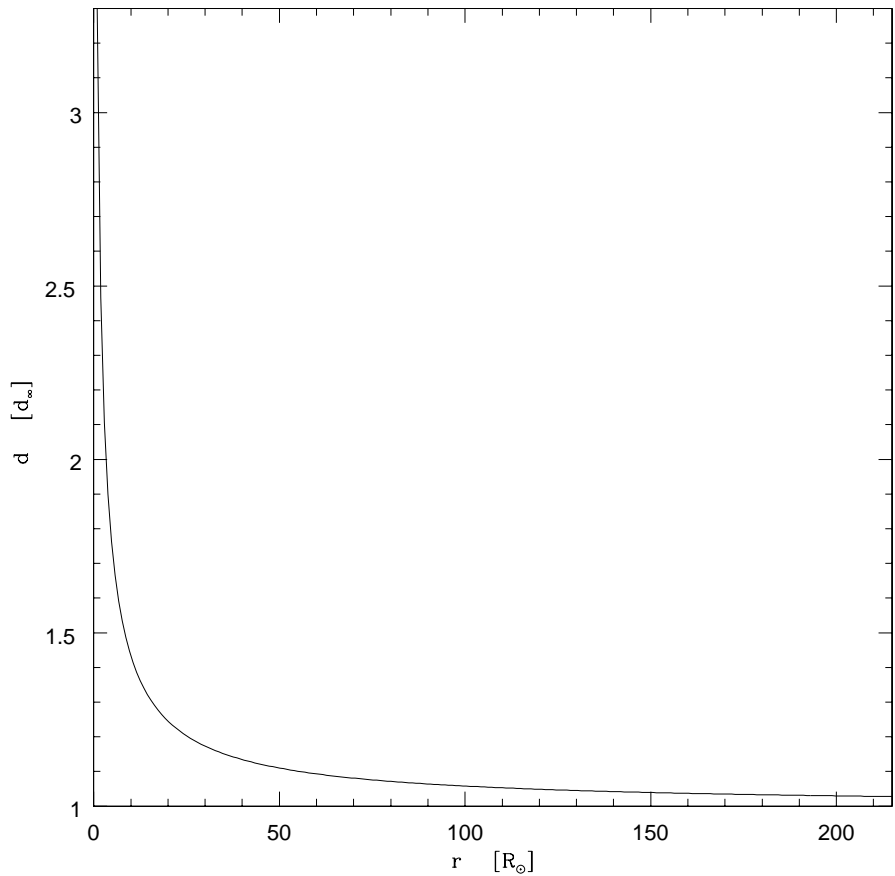}}
\caption{The density as a function of distance to the Sun, in the 
isothermal model with $\sqrt{\vec{v}^2}$ = 270 km/s, if the Sun 
were not moving with respect to the halo ($v_{\odot}=0$).  The 
increase in density at small $r$, compared to $d_\infty$, is the 
effect of the Sun's gravity.  At the Earth's distance the increase 
is approximately $3\%$.}
\label{fig:9}
\end{figure}

The effect of the Sun's gravity on the event rate in WIMP searches is 
largely due to the change in the density of dark matter.  Moreover the 
signal from axion to photon conversion in the cavity detector of galactic 
halo axions is simply proportional to the dark matter density.  So let us
calculate the density $d(r,\theta)$ where $r$ is the distance to the Sun 
and $\theta$ is the angle between $\vec{r}$ and $\vec{v}_{\odot}.$  We 
could integrate Eq.~(\ref{eq:5.9}) over $\vec{v}.$  However, it is
easier to integrate
\beq
d(r,\theta)=\int d^{3}v_\infty f(\vec{v}_\infty)d(\vec{r},\vec{v}_\infty)
\label{eq:5.11}
\eeq
where $d(\vec{r},\vec{v}_\infty)$ is the density due to a single
flow of initial velocity $\vec{v}_\infty,$ Eq.~(\ref{eq:2.11}).
This yields
\bea
d(r,\theta)= {d_\infty \over \pi^{3 \over 2}}
\int_{0}^{\infty}u^{2}du\int_{-1}^{+1}ds
\int_{0}^{2\pi}d\varphi~ &~&  
\exp[-(u^{2}+u_{\odot}^{2}+2uu_{\odot}s)] \cdot\nonumber\\
 \cdot~&{1 \over 2}&\left(\sqrt{1+{4 \over X}}
+ {1 \over \sqrt{1+ {4 \over X}}}\right)
\label{eq:5.12}
\eea
where $u_{\odot}=\frac{v_{\odot}}{\sigma}$,
\beq
X = \frac{4}{y} = \frac{r}{a_\sigma}~u^2
(1-s\cos\theta-\sqrt{1-s^{2}}\sin\theta\cos\varphi)
\label{eq:5.13}
\eeq
and $a_\sigma \equiv {GM_\odot \over \sigma^2}$. The integral in 
Eq. (\ref{eq:5.12}) was evaluated numerically.  Fig.~\ref{fig:9} 
shows $d(r)$ in case the Sun were at rest with respect to the halo 
($v_\odot =0$).  At the Earth's distance from the Sun, the density 
increase due to the Sun's gravity is approximately 3\%.

\begin{figure}
\epsfxsize=5in
\centerline{\epsfbox{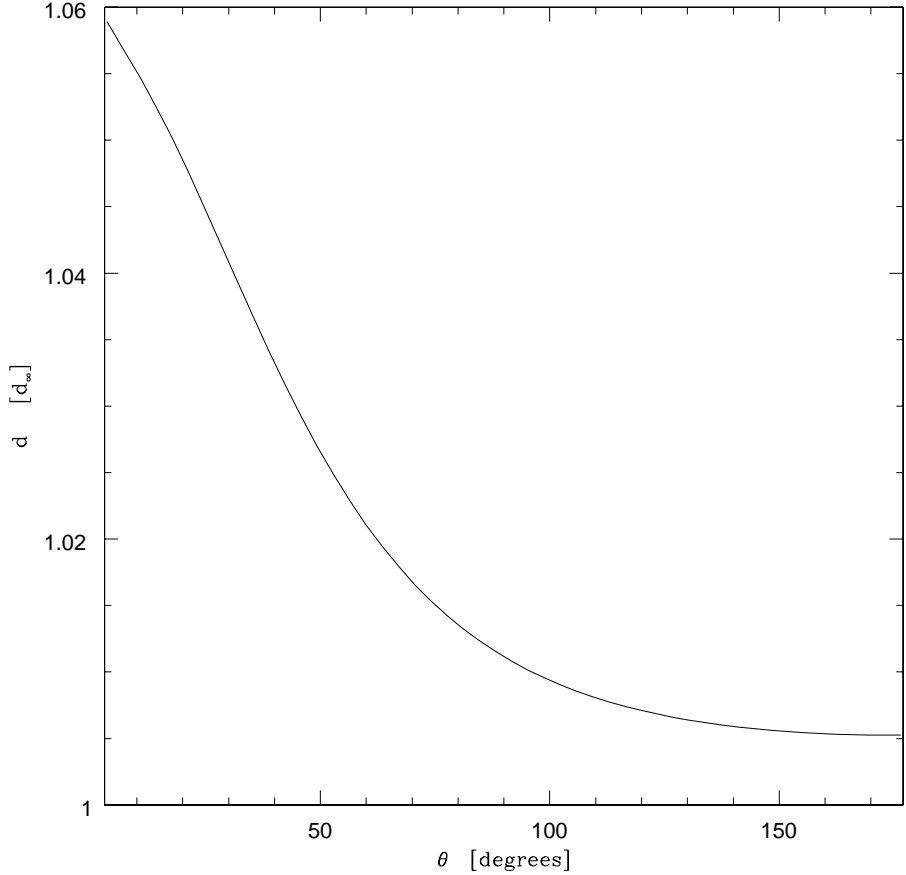}}
\caption{ 
The density as a function of polar angle $\theta$ at 
$r=1$~A.U.~in the isothermal model 
with $\sqrt{\vec{v}^2}$ = 270 km/s 
and $v_{\odot}= $ 220 km/s.  $\theta$ is relative to the 
direction of $-\vec{v}_{\odot}$.  The Earth's orbit 
is inclined at approximately $60^{\circ}$ relative to 
$\vec{v}_{\odot}$.  The annual modulation due to the Sun's 
gravity can be read off by comparing $\theta\sim~60^{\circ}$ 
(March 3) and $\theta\sim~120^{\circ}$ (September 2). It is 
approximately 0.7\% in amplitude.}
\label{fig:10}
\end{figure}

Fig.~\ref{fig:10} shows $d(r,\theta)$ as a function of $\theta$
for $v_{\odot}=\sigma=$ 220~km/s and
$r=1.5\times~10^{13}$~cm, the mean Earth--Sun distance.  At the Earth's
distance from the Sun, the density increase is approximately 0.5\%
upstream ($\theta=\pi$) and 6\% downstream ($\theta=0$).   
The large density downstream 
is the spike caustic but spread out because of the large velocity 
dispersion. The Earth's
orbit is inclined at approximately $60^{\circ}$ relative to 
$\vec{v}_\odot$, with the Earth being most closely downstream of the Sun
on March 3.  Fig. \ref{fig:10} shows that the density variation between
September 2 ($\theta=120^{\circ}$) and March 3 ($\theta=60^{\circ}$) is 
of order 1.4\%.  This is consistent with the earlier calculation of Griest
\cite{gri}.

\section{Conclusions}
\label{sec:conclusion}

We analyzed the effect of the Sun's gravity on a flow of collisionless
dark matter through the solar neighborhood.  We were mainly concerned with
the case of a cold flow, i.e. a flow with zero velocity dispersion.  The
cold flow results can be generalized to arbitrary flows by integrating
over the initial $\vec{v}_\infty$ velocity distribution.  We neglected the
effect of the Earth's gravity, but this is indeed small. The ratio of
the Sun and Earth gravity effects is of order
$\frac{M_{\oplus}}{R_{\oplus}}\frac{r}{M_{\odot}}\simeq~0.07$ where
$M_{\oplus}$ is the Earth's mass, $R_{\oplus}$ its radius, and $r$ is the
mean Earth--Sun distance.

In section II, we obtained formulas for the density and velocity 
distributions of cold flows with axial symmetry.  In section III, these 
formulas were applied to a cold uniform flow past a point mass.  In 
that case the number of flows $n = 2$ everywhere.  There is a caustic
line, which we called the 'spike', downstream of the Sun.  It is located
on the positive $\hat{z}$--axis where $\hat{z}$ is the direction of the
incoming flow ($\vec{v}_\infty = v_\infty \hat{z}$).  The profile of the 
spike is given in Eq. (\ref{eq:2.12}).  The densities of the two flows 
everywhere are given by Eqs. (\ref{eq:2.9}), the velocity distributions 
by Eqs. (\ref{eq:2.10}).  These are exact results.

In section III, we took account of the Sun's finite radius.  We assumed
that the solar mass distribution is spherically symmetric.  Many of the
results derived in section III are still valid.  In particular, the spike
caustic remains and its profile is still given by Eq. (\ref{eq:2.12}) for
$z$ large enough that the trajectories there did not go through the Sun in
the past.  A new caustic appears which we called the 'skirt'.  It is
located near the cone of axis $\hat{z}$ and opening angle $\Theta_{\rm
max}$ where $\Theta_{\rm max}$ is the maximum scattering angle.  Inside
this cone $n=3$, whereas $n=1$ outside the cone.  The density enhancement
associated with the skirt falls off as ${1 \over r^2}$ where $r$ is the
distance to the Sun, and is rather feeble on Earth even at the moment of
skirt crossing.  However, the profile of the density near the skirt
depends on the mass distribution inside the Sun.  In principle, by
monitoring the densities of the various dark matter flows as we orbit the
Sun, the mass distribution of the Sun can be investigated.  The density
profile near a skirt was obtained in terms of quantities $H$, $b_c$ and
$\Theta_{\rm max}$, which we defined and computed as a function of
$v_\infty$ for a realistic model of the solar mass distribution.

In section IV, we applied our methods to the single flow of dark matter
predicted by the isothermal model.  That flow has direction
$-\hat{\phi}_G$ opposite to the Sun's motion in the Galaxy and velocity
dispersion $\sqrt{\vec{v}^2} \simeq 270$ km/s.  We derived a formula which
gives the velocity distribution at any point in the neighborhood of the
Sun in terms of an arbitrary initial velocity distribution
$f(\vec{v}_\infty)$. We also computed the density as a function of
position relative to the Sun.  There is a spike caustic downstream of the
Sun but it is spread out because of the large velocity dispersion.  In the
isothermal model the density is largest (smallest) near March 3 (September
2).  The amplitude of this density modulation is of order 0.7\%.

\section{Acknowledgments}

P.S. would like to thank Doug Eardly for stimulating discussions on this
topic several years ago.  He is also grateful to the Aspen Center for
Physics for its hospitality while working on this paper.  This work was
supported in part by the U.S. Department of Energy under grant
DE-FG02-97ER41029.

\newpage

\newpage




\end{document}